\documentclass[reprint,amsmath,amssymb,aps]{revtex4-2}
\pdfoutput=1

\usepackage{graphicx}
\usepackage{dcolumn}
\usepackage{bm}
\usepackage{braket}
\usepackage[lined,boxruled]{algorithm2e}
\newtheorem{theorem}{Theorem}

\begin{document}

\preprint{APS/123-QED}

\title{$O(N^3)$ Measurement Cost for Variational Quantum Eigensolver \\ on Molecular Hamiltonians}

\author{Pranav Gokhale}
 \email{pranavgokhale@uchicago.edu}
\author{Frederic T. Chong}
\affiliation{
 Department of Computer Science, University of Chicago
}

\collaboration{EPiQC Expedition}

\date{\today}

\begin{abstract}
Variational Quantum Eigensolver (VQE) is a promising algorithm for near-term quantum machines. It can be used to estimate the ground state energy of a molecule by performing separate measurements of $O(N^4)$ terms. Several recent papers observed that this scaling may be reducible to $O(N^3)$ by partitioning the terms into linear-sized commuting families that can be measured simultaneously. We confirm these empirical observations by studying the MIN-COMMUTING-PARTITION problem at the level of the fermionic Hamiltonian and its encoding into qubits. Moreover, we provide a fast, pre-computable procedure for creating linearly-sized commuting partitions by solving a round-robin scheduling problem via flow networks.
\end{abstract}

\maketitle


\section{Background}
Variational Quantum Eigensolver (VQE) \cite{peruzzo2014variational} is a quantum algorithm that is a leading contender, if not the top contender, for demonstrating a practical quantum advantage on near-term machines. Unlike traditional quantum algorithms, which have extremely high quantum requirements in terms of gate counts and qubit lifetimes, VQE is feasible with modest quantum resources that are already available on current quantum computers. It attains a lower quantum resource cost in part by structuring computation over a large number of subproblems, each of which can be performed on a quantum computer with modest capabilities.

While the low quantum resource requirements per subproblem are appealing, the number of subproblems is an issue for practical application of VQE. Consider molecular ground state estimation, a classically-hard problem that is considered the canonical application of VQE. Within the framework of VQE, molecular energy estimation is performed by applying linearity of expectation to the Hamiltonian $H$, an observable that captures a molecule's energy configuration. Under the second quantization and expressed in fermionic form, we have \cite{mcardle2018quantum}:

\begin{equation}
    H = \sum_{p=1}^N \sum_{q=1}^N h_{pq} a_p^{\dagger} a_q + \sum_{p=1}^N \sum_{q=1}^N \sum_{r=1}^N \sum_{s=1}^N h_{pqrs} a^{\dagger}_p a^{\dagger}_q a_r a_s
\label{eq:second_quantization}
\end{equation}

Applying linearity of expectation, we see that measuring $\braket{H}$ reduces to measuring $\braket{a_p^{\dagger} a_q}$ and $\braket{a^{\dagger}_p a^{\dagger}_q a_r a_s}$. Each of these $O(N^4)$ terms is transformed via fermion-to-qubit encoding into a sum over a constant number of Pauli strings ($N$-fold tensor product of Pauli matrices). Measurement of each of these resulting $O(N^4)$ Pauli strings constitutes a subproblem. Although the measurement for each subproblem is simple, requiring only single-qubit rotations, the $O(N^4)$ scaling of subproblems poses a practical challenge towards applying VQE to molecules of interest such as caffeine and cholesterol, which appear to require $N$ numbering hundreds of qubits \cite{aspuru2005simulated}.

Recently however, several research groups observed that this $O(N^4)$ scaling may be reducible to $O(N^3)$ \cite{gokhale2019minimizing, jena2019pauli, yen2019measuring, izmaylov2019unitary, crawford2019efficient}. The core principle underlying these papers is that commuting Pauli strings can be measured simultaneously. The $O(N^4) \to O(N^3)$ improvement is conjectured based on extrapolation of results across a range of molecules.

Here, we confirm this observation of linearly-reduced measurement cost for molecular Hamiltonians encoded under Jordan-Wigner---the most widely used encoding \cite{huggins2019efficient}. Our general approach is to demonstrate that the molecular Hamiltonians can always be partitioned into pairwise-commuting families where each family contains $O(N)$ terms. Since the terms in each such family can be measured simultaneously, this constitutes our reduction in the measurement cost of VQE from $O(N^4)$ to $O(N^3)$.

In addition to proving the existence of such a partition, we explicitly demonstrate how to construct it. Our construction is efficient, computable in $O(N^5 \log N)$ time. Moreover, the construction is independent of the specific molecular Hamiltonian of interest and instead only depends on $N$. This means that the partitioning can be pre-computed once for each $N$. The efficiency of our approach is critical. In contrast, proposals for simultaneous measurement in the recent prior work have involved algorithms with runtimes as high as $O(N^{12})$, which may be slow enough to undermine the advantage of simultaneous measurement.

\section{Prior Work}

The empirical results in \cite{gokhale2019minimizing, jena2019pauli, yen2019measuring, izmaylov2019unitary, crawford2019efficient} all suggest that the advantage due to simultaneous measurement appears to increase for larger molecules. The specificity of this claim varies across the papers---\cite{gokhale2019minimizing} explicitly extrapolates linear scaling for molecular Hamiltonians over a range of encodings, molecules, and active space sizes; \cite{jena2019pauli} formulates it as an explicit conjecture for ``almost all'' sets of Pauli strings; \cite{yen2019measuring, izmaylov2019unitary} observes this scaling via least-squares fitting for molecular Hamiltonians under the Jordan-Wigner and Bravyi-Kitaev qubit encodings; and \cite{crawford2019efficient} makes note of increasing partition size with increasing $N$.

Moreover, \cite[Section 5.1]{gokhale2019minimizing} provides two encouraging examples of an asymptotic gain from simultaneous measurement for \textit{specific} types of contrived Hamiltonians. First, it is observed that simultaneous measurement can yield an exponential gain: the $2^N$ Pauli strings with the same underlying measurement basis across all qubits can be simultaneously measured with a single measurement. Second, in the case of measuring all $4^N$ Pauli strings on $N$ qubits, a square root ($2^N$) reduction is achievable by Mutually Unbiased Bases.

However, as suggested by \cite[Appendix A]{gokhale2019minimizing}, an asymptotic gain from simultaneous measurement is not guaranteed. For example, consider the set of $2N$ Pauli strings matching the pattern Z*(X$|$Y)I*, where * matches 0 or more occurrences and  $|$ is a Boolean OR. For example, for $N = 3$, we have $[XII, YII, ZXI, ZYI, ZZX, ZZY]$. It can be shown that none of the pairs in this set commute. Thus, simultaneous measurement offers no advantage for this set of Pauli strings. More generally, we see that simultaneous measurement does not automatically confer any advantage.

During the preparation of this manuscript, we became aware of very recent work by \cite{zhao2019measurement} that also proves the $O(N^3)$ measurement cost for molecular Hamiltonians. Their work approaches the problem via Majorana operators, which leads to a proof agnostic of the underlying fermion-to-qubit encoding.




\section{Commutativity of Index-Disjoint Terms}
Our top-level goal is to partition the molecular Hamiltonian into commuting families, such that the number of partitions is minimized. This problem is termed MIN-COMMUTING-PARTITION and is NP-Hard in general \cite{gokhale2019minimizing}. We instead seek to \textit{approximate} a good partitioning. Our approach is to address this problem at the level of the fermionic Hamiltonian in Equation~\ref{eq:second_quantization}. By contrast, past work, except for \cite[Section 6]{gokhale2019minimizing}, has focused on this problem at the qubit Hamiltonian stage, after the fermionic Hamiltonian has been encoded into a summation over Pauli strings.

We focus on the $O(N^4)$ terms with $p \neq q \neq r \neq s$ in the second sum of Equation~\ref{eq:second_quantization}, because these terms are asymptotically dominant; the number of other terms is only $O(N^3)$. Without loss of generality, let us suppose that $p > q > r > s$, and likewise $i > j > k > l$. We denote the set of Pauli strings in the Jordan-Wigner encoding of $a^{\dagger}_p a^{\dagger}_q a_r a_s$ as $\{a^{\dagger}_p a^{\dagger}_q a_r a_s\}_\text{JW}$.

Our core observation is that if two $a^{\dagger} a^{\dagger} a a$ terms have disjoint indices, then the terms in their qubit encodings commute. In particular:

\begin{theorem} \label{thrm:commutator}
If $\{p, q, r, s\} \cap \{i, j, k, l\} = \emptyset$, then
$$
[\{a^{\dagger}_p a^{\dagger}_q a_r a_s\}_\text{JW},\, \{a^{\dagger}_i a^{\dagger}_j a_k a_l\}_\text{JW}] = 0
$$
where the commutator is taken to apply between all pairs of elements between the two sets.
\end{theorem}

\begin{figure}[h]
    \centering
    \includegraphics[width=0.45\textwidth]{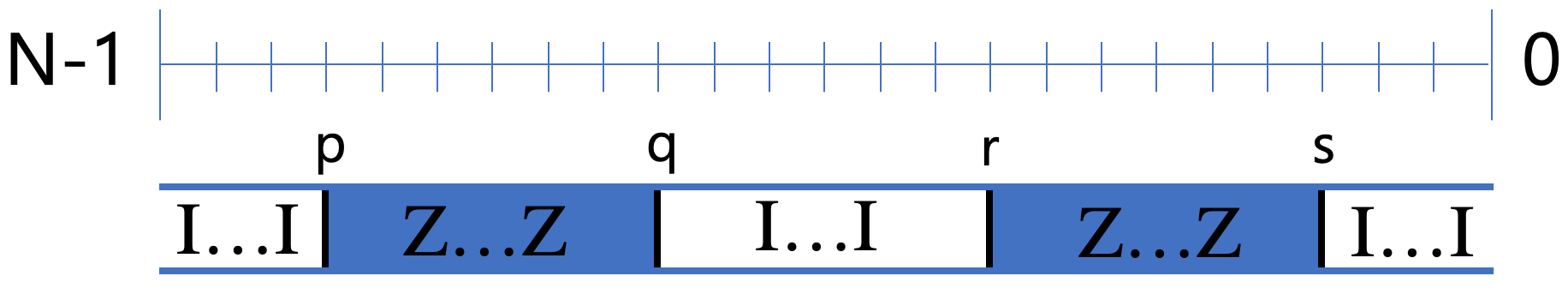}
    \caption{Pictorial representation of the Jordan-Wigner encoding of $a^{\dagger}_p a^{\dagger}_q a_r a_s$. Repeating $Z$'s span the blue rectangles between $p$ and $q$ and between $r$ and $s$. The other three ranges have repeating $I$'s. At indices $p$, $q$, $r$, and $s$, which are denoted by the black vertical bars between the blue and white rectangles, we can have either $X$ or $Y$. Thus, there are $2^4$ = 16 Pauli strings involved in the Jordan-Wigner encoding.}
    \label{fig:pqrs_rectangle}
\end{figure}

Theorem~\ref{thrm:commutator} can be verified by inspecting the form of the Pauli string terms in $\{a^{\dagger} a^{\dagger} a a\}_\text{JW}$. Under the Jordan-Wigner encoding \cite{jordan1928pauli}, we perform the transformations:
$$a_p \rightarrow \frac{X_p + i Y_p}{2} Z_{p-1} ... Z_0, \quad a_p^{\dagger} \rightarrow \frac{X_p - i Y_p}{2} Z_{p-1} ... Z_0 $$
Carrying out the transformation for $a^{\dagger}_p a^{\dagger}_q a_r a_s$ yields the 16 Pauli strings matching the regular expression:
\begin{equation*}
    (X_p|Y_p)\overline{Z}_{p:q}(X_q | Y_q) (X_r | Y_r)\overline{Z}_{r:s}(X_s | Y_s)
\end{equation*}
where $\overline{Z}_{p:q}$ denotes $Z$ on each index between $p$ and $q$, \textit{exclusive} of endpoints. Figure~\ref{fig:pqrs_rectangle} shows this pattern as a pictorial representation: the repeating $Z$'s are blue rectangles and the $\{p,q,r,s\}$ indices are the black vertical bars demarcating the blue and white rectangles.

\begin{figure}[h]
    \centering
    \includegraphics[width=0.45\textwidth]{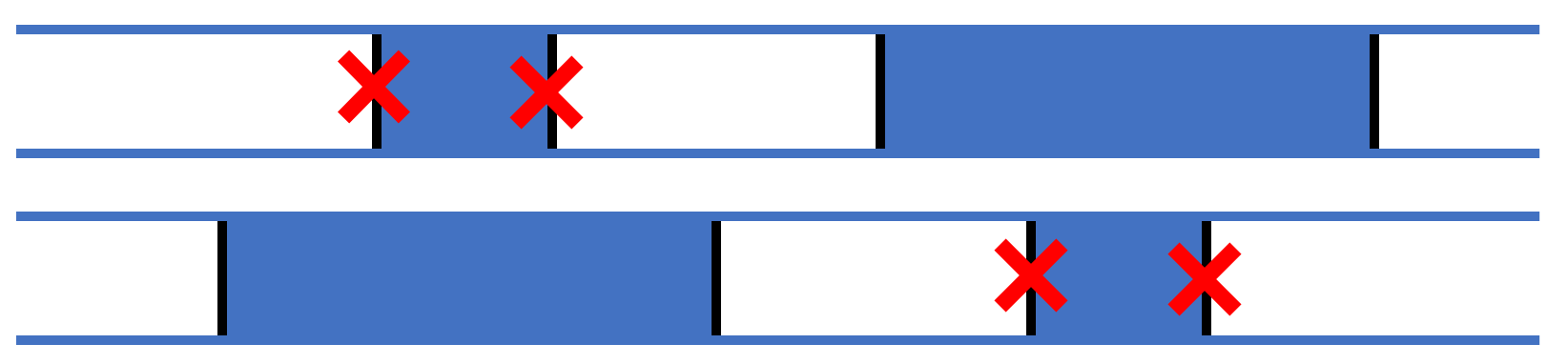}
    \caption{Pictorial representation of the commutation on each index between two $\{a^{\dagger} a^{\dagger} a a\}_\text{JW}$ rectangles. All indices commute except possibly the 8 indices with black bars---these indices anti-commute when the black bar ($X$ or $Y$) is vertically aligned with a blue rectangle $Z$. In this example, the there are an even (4) number of anti-commuting terms, so the two patterns commute.}
    \label{fig:rectangle_commutativity}
\end{figure}

To evaluate the commutativity between a term in $\{a^{\dagger}_p a^{\dagger}_q a_r a_s\}_\text{JW}$ and a term in $\{a^{\dagger}_i a^{\dagger}_j a_k a_l\}_\text{JW}$, we simply need to count the number of indices that anti-commute, as explained in \cite[Section 3]{gokhale2019minimizing}. If the number of anti-commuting indices is even, then the two Pauli strings commute. For all indices other than $p, q, r, s, i, j, k, l$, the Pauli matrices at the indices commute, because $[I, I] = [I, Z] = [Z, I] = [Z, Z]$ = 0. On the remaining 8 indices, the commutation depends on whether the $(X|Y)$ is matched to an $I$ (commutes) or $Z$ (anti-commutes). Figure~\ref{fig:rectangle_commutativity} depicts this: when one of the black bars ($X|Y$) is vertically aligned with a blue rectangle ($Z$), the index does not commute, as marked by the red cross. When the black bar is vertically aligned with a white rectangle ($I$), the index commutes.

\begin{figure}[h]
    \centering
    \includegraphics[width=0.48\textwidth]{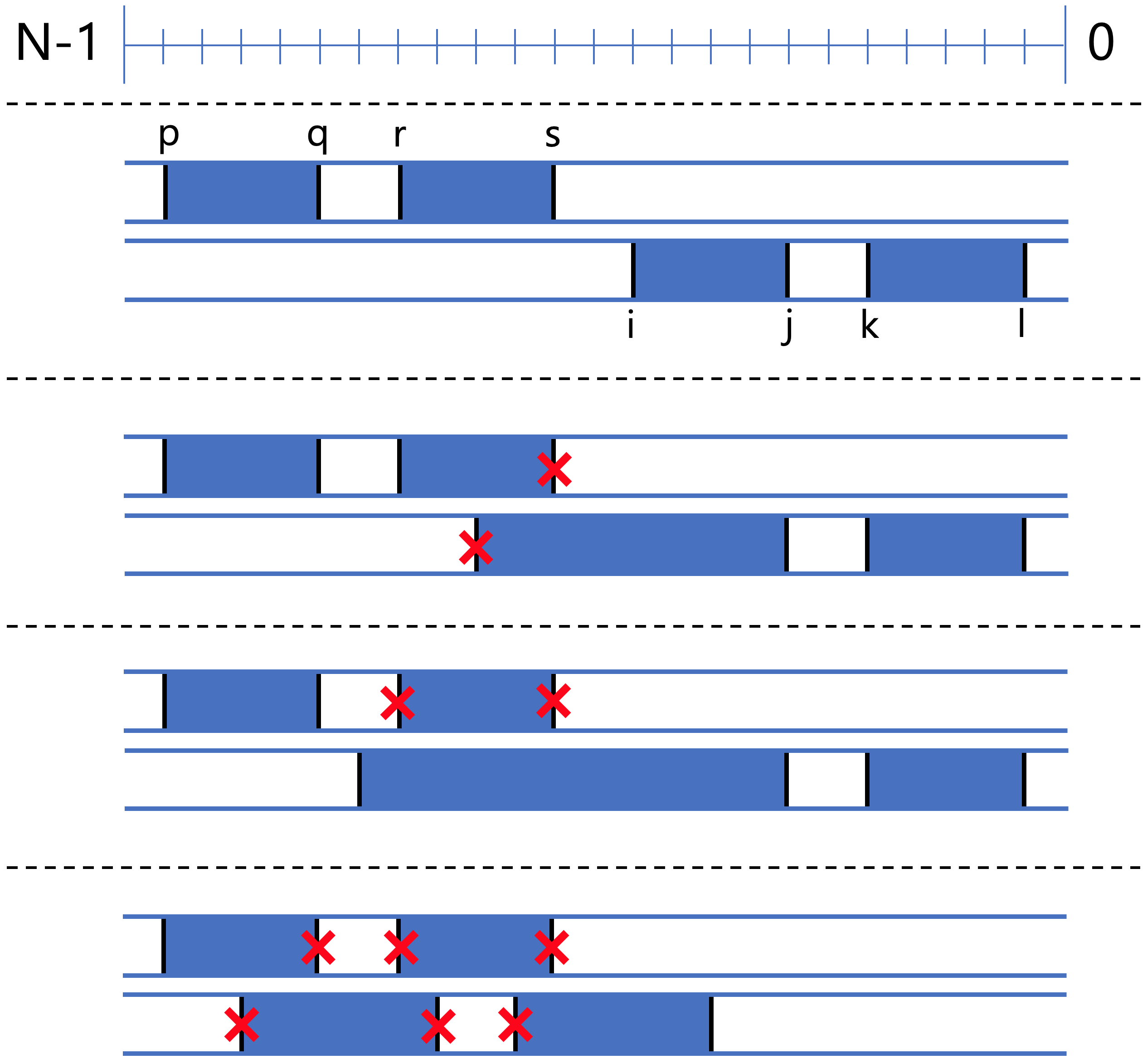}
    \caption{Four representative examples illustrating why $\{a^{\dagger}_p a^{\dagger}_q a_r a_s\}_\text{JW}$ and $\{a^{\dagger}_i a^{\dagger}_j a_k a_l\}_\text{JW}$ terms always commute (have an even number of anti-commuting indices) when $\{p,q,r,s\} \cap \{i,j,k,l\} = \emptyset$. At the top, no black bars align with blue rectangles, so there are 0 anti-commuting indices. Below, $r > i > s > j$, so there are 2 anti-commuting indices: $i$ and $s$. Below that, observe that sliding the $i$ endpoint into the interval between $q$ and $r$ does not change the parity of the number of anti-commuting indices. The bottom example shows a case with the maximal number of anti-commuting indices, 6.}
    \label{fig:four_cases}
\end{figure}

The commutativity between $\{a^{\dagger}_p a^{\dagger}_q a_r a_s\}_\text{JW}$ and $\{a^{\dagger}_i a^{\dagger}_j a_k a_l\}_\text{JW}$ terms can be verified by considering all possible interleaved orderings of the 8 indices, subject to the constraint that $p > q > r > s$ and $i > j > k > l$. There are $\binom{8}{4} = 70$ such cases that can be explicitly checked (or 35 cases, accounting for symmetry) to prove Theorem~\ref{thrm:commutator}. Figure~\ref{fig:four_cases} demonstrates four representative cases, which provide useful intuition for the general case. In particular, when sliding one of the $\{p,q,r,s\}$ indices while keeping $\{i,j,k,l\}$ fixed, the parity of the number of anti-commuting indices is invariant. Thus, this parity is always even, and two $\{a^{\dagger}_p a^{\dagger}_q a_r a_s\}_\text{JW}$ and $\{a^{\dagger}_i a^{\dagger}_j a_k a_l\}_\text{JW}$ terms with disjoint indices always commute, as claimed in Theorem~\ref{thrm:commutator}.

\section{Existence of Linearly-Sized Partitions}

Consider the set of Pauli strings contained in
$$\{a^{\dagger}_N a^{\dagger}_{N-1} a_{N-2} a_{N-3}\}_{\text{JW}} \, \cup \, \{a^{\dagger}_8 a^{\dagger}_7 a_6 a_5\}_{\text{JW}} \, \cup \, ... \, \cup \, \{a^{\dagger}_4 a^{\dagger}_3 a_2 a_1\}_{\text{JW}}$$
for $N$ divisible by 4. There are $16 \frac{N}{4} = 4N \in O(N)$ Pauli strings in this set. However, since the indices are disjoint, Pauli strings from each of the $\frac{N}{4}$ subsets can be measured simultaneously by Theorem~\ref{thrm:commutator}. In particular, the Pauli strings can be partitioned into $16 \in O(1)$ measurement families. In fact, they can even be partitioned into just $2$ measurement families by noting that the MIN-COMMUTING-PARTITION within each $\{a^{\dagger}_p a^{\dagger}_q a_r a_s\}_\text{JW}$ term is 2, as described in \cite[Section 6]{gokhale2019minimizing}.

A natural question is whether \textbf{all} $\binom{N}{4}$ $p > q > r > s$ terms in Equation~\ref{eq:second_quantization} can be partitioned in such a fashion---if so, then this constitutes a partitioning of the $O(N^4)$ terms into $\binom{N}{4} / \frac{N}{4} = \binom{N-1}{3} \in O(N^3)$ commuting families. Intuitively, this is the same problem as trying to schedule a round-robin tournament of $N$ players with 4 players-per-game into $\binom{N-1}{3}$ rounds. We can think of each index as a player, and 4-player games can be scheduled simultaneously if they don't share players. Equivalently, these problems can be bijected to a graph theory problem: does the 4-uniform complete hypergraph on $N$ vertices admit a 1-factorization?

The answer to all of these questions is affirmative, per Baranyai's Theorem \cite{baranyai1974factorization}. In our case, it means that for $N$ divisible by 4, the $\binom{N}{4} \in O(N^4)$ terms can be partitioned into $\binom{N-1}{3} \in O(N^3)$ sets, such that the $\frac{N}{4}$ terms within each set have disjoint indices. Table~I demonstrates such a partitioning for $N=8$ qubits. Each of the $\binom{8-1}{3} = 35$ rows has two fermionic terms with disjoint indices---thus, their corresponding Jordan-Wigner qubit encodings can be measured simultaneously.

\renewcommand{\arraystretch}{1.2} 
\begin{table}[h]\label{tab:8qubitpartition}
\begin{tabular}{c}
\hline
$a^{\dagger}_7 a^{\dagger}_5 a_3 a_0$ \quad $a^{\dagger}_6 a^{\dagger}_4 a_2 a_1$ \\ \hline 
$a^{\dagger}_6 a^{\dagger}_5 a_3 a_0$ \quad $a^{\dagger}_7 a^{\dagger}_4 a_2 a_1$ \\ \hline 
$a^{\dagger}_7 a^{\dagger}_6 a_3 a_0$ \quad $a^{\dagger}_5 a^{\dagger}_4 a_2 a_1$ \\ \hline 
$a^{\dagger}_7 a^{\dagger}_4 a_3 a_0$ \quad $a^{\dagger}_6 a^{\dagger}_5 a_2 a_1$ \\ \hline 
$a^{\dagger}_7 a^{\dagger}_5 a_4 a_0$ \quad $a^{\dagger}_6 a^{\dagger}_3 a_2 a_1$ \\ \hline 
$a^{\dagger}_6 a^{\dagger}_4 a_3 a_0$ \quad $a^{\dagger}_7 a^{\dagger}_5 a_2 a_1$ \\ \hline 
$a^{\dagger}_6 a^{\dagger}_5 a_4 a_0$ \quad $a^{\dagger}_7 a^{\dagger}_3 a_2 a_1$ \\ \hline 
$a^{\dagger}_7 a^{\dagger}_6 a_4 a_0$ \quad $a^{\dagger}_5 a^{\dagger}_3 a_2 a_1$ \\ \hline 
$a^{\dagger}_5 a^{\dagger}_4 a_3 a_0$ \quad $a^{\dagger}_7 a^{\dagger}_6 a_2 a_1$ \\ \hline 
$a^{\dagger}_7 a^{\dagger}_6 a_5 a_0$ \quad $a^{\dagger}_4 a^{\dagger}_3 a_2 a_1$ \\ \hline 
$a^{\dagger}_7 a^{\dagger}_5 a_1 a_0$ \quad $a^{\dagger}_6 a^{\dagger}_4 a_3 a_2$ \\ \hline 
$a^{\dagger}_7 a^{\dagger}_5 a_2 a_0$ \quad $a^{\dagger}_6 a^{\dagger}_4 a_3 a_1$ \\ \hline 
$a^{\dagger}_6 a^{\dagger}_5 a_1 a_0$ \quad $a^{\dagger}_7 a^{\dagger}_4 a_3 a_2$ \\ \hline 
$a^{\dagger}_6 a^{\dagger}_5 a_2 a_0$ \quad $a^{\dagger}_7 a^{\dagger}_4 a_3 a_1$ \\ \hline 
$a^{\dagger}_7 a^{\dagger}_6 a_1 a_0$ \quad $a^{\dagger}_5 a^{\dagger}_4 a_3 a_2$ \\ \hline 
$a^{\dagger}_7 a^{\dagger}_4 a_1 a_0$ \quad $a^{\dagger}_6 a^{\dagger}_5 a_3 a_2$ \\ \hline 
$a^{\dagger}_7 a^{\dagger}_6 a_2 a_0$ \quad $a^{\dagger}_5 a^{\dagger}_4 a_3 a_1$ \\ \hline 
$a^{\dagger}_7 a^{\dagger}_3 a_1 a_0$ \quad $a^{\dagger}_6 a^{\dagger}_5 a_4 a_2$ \\ \hline 
$a^{\dagger}_7 a^{\dagger}_4 a_2 a_0$ \quad $a^{\dagger}_6 a^{\dagger}_5 a_3 a_1$ \\ \hline 
$a^{\dagger}_6 a^{\dagger}_4 a_1 a_0$ \quad $a^{\dagger}_7 a^{\dagger}_5 a_3 a_2$ \\ \hline 
$a^{\dagger}_6 a^{\dagger}_3 a_1 a_0$ \quad $a^{\dagger}_7 a^{\dagger}_5 a_4 a_2$ \\ \hline 
$a^{\dagger}_7 a^{\dagger}_3 a_2 a_0$ \quad $a^{\dagger}_6 a^{\dagger}_5 a_4 a_1$ \\ \hline 
$a^{\dagger}_5 a^{\dagger}_3 a_1 a_0$ \quad $a^{\dagger}_7 a^{\dagger}_6 a_4 a_2$ \\ \hline 
$a^{\dagger}_6 a^{\dagger}_4 a_2 a_0$ \quad $a^{\dagger}_7 a^{\dagger}_5 a_3 a_1$ \\ \hline 
$a^{\dagger}_5 a^{\dagger}_4 a_1 a_0$ \quad $a^{\dagger}_7 a^{\dagger}_6 a_3 a_2$ \\ \hline 
$a^{\dagger}_4 a^{\dagger}_3 a_1 a_0$ \quad $a^{\dagger}_7 a^{\dagger}_6 a_5 a_2$ \\ \hline 
$a^{\dagger}_6 a^{\dagger}_3 a_2 a_0$ \quad $a^{\dagger}_7 a^{\dagger}_5 a_4 a_1$ \\ \hline 
$a^{\dagger}_7 a^{\dagger}_2 a_1 a_0$ \quad $a^{\dagger}_6 a^{\dagger}_5 a_4 a_3$ \\ \hline 
$a^{\dagger}_5 a^{\dagger}_3 a_2 a_0$ \quad $a^{\dagger}_7 a^{\dagger}_6 a_4 a_1$ \\ \hline 
$a^{\dagger}_6 a^{\dagger}_2 a_1 a_0$ \quad $a^{\dagger}_7 a^{\dagger}_5 a_4 a_3$ \\ \hline 
$a^{\dagger}_5 a^{\dagger}_2 a_1 a_0$ \quad $a^{\dagger}_7 a^{\dagger}_6 a_4 a_3$ \\ \hline 
$a^{\dagger}_5 a^{\dagger}_4 a_2 a_0$ \quad $a^{\dagger}_7 a^{\dagger}_6 a_3 a_1$ \\ \hline 
$a^{\dagger}_4 a^{\dagger}_2 a_1 a_0$ \quad $a^{\dagger}_7 a^{\dagger}_6 a_5 a_3$ \\ \hline 
$a^{\dagger}_4 a^{\dagger}_3 a_2 a_0$ \quad $a^{\dagger}_7 a^{\dagger}_6 a_5 a_1$ \\ \hline 
$a^{\dagger}_3 a^{\dagger}_2 a_1 a_0$ \quad $a^{\dagger}_7 a^{\dagger}_6 a_5 a_4$ \\ \hline 
\end{tabular}
\caption{Partitioning of $\binom{N=8}{4} = 70$ $a^{\dagger}_p a^{\dagger}_q a_r a_s$ terms into $\binom{N-1=7}{3} = 35$ subsets, with disjoint indices between the two terms in each subset. Such a partitioning is guaranteed to exist for all $N$ divisible by 4, per Baranyai's Theorem \cite{baranyai1974factorization}.}
\end{table}

\section{Construction of Linearly-Sized Partitions}
Prior literature refers to Baranyai's original proof as either being non-constructive \cite{bailey2010hamiltonian, tammapplications} or providing an exponential-time construction \cite{deo2002one} (prior literature varies in what exactly is considered Baranyai's proof). In order for Baranyai's proof to be useful to us, we need a fast polynomial-time algorithm for partitioning the $\binom{N}{4}$ subsets of $N$ into $\binom{N-1}{3}$ groups, each containing $N/4$ disjoint subsets. Fortunately, due to later work by \cite{brouwer1979uniform}, a proof was provided that leads to an efficient construction \cite{weisstein2019baranyai}. The proof is based on maximum flows in network flow graphs.

We refer readers to \cite{bartlett2015lecture} for a lucid explanation and to \cite{eisenstatstackoverflow} for an implementation in code. This implementation was used to generate Table I. The pseudocode is given in Algorithm~\ref{alg:baranyai}. An outer loop is called $N$ times, and each iteration solves for maximum flow on a network with $O(N^3)$ vertices and $O(N^4)$ directed edges. Since the maximum flow in the proof construction has a value of $O(N^3)$, solving for it with the Ford-Fulkerson algorithm would incur a cost of $O(N^7)$ per loop iteration \cite{ford1954maximal}. However, due to work on flow-rounding \cite{cohen1995approximate, madry2013navigating, lee2013new, kang2015flow}, this runtime is reduced to $O(N^4 \log{N})$. This is because for each flow network, a \textit{fractional} solution is known that can be rounded to an integral solution faster than computing an integral solution from scratch. Thus, the total runtime of the Baranyai constructive proof is $O(N^5 \log{N})$.

\begin{algorithm}
\SetAlgoLined
\SetKwInOut{Input}{input}\SetKwInOut{Output}{output}
\Input{ $N$}
\Output{ $\binom{N-1}{3}$ sets where each set contains $\frac{N}{4}$ disjoint size-4 subsets, and no term is repeated}
\BlankLine
\For{$i \in [1, 2, ..., N]$}{
Create flow network with two layers: $O(N^3)$ partition nodes and $O(N^3)$ subset nodes\;
Set capacities for $O(N^4)$ edges per \cite{brouwer1979uniform, bartlett2015lecture} construction\;
Set fractional maxflow of value $\binom{N-1}{3}$, saturating nodes out of source and into destination\;
Round fractional maxflow into an integral maxflow\;
Update subset nodes based on integral maxflow\;
}
Return schedule of $\binom{N}{4}$ subsets, based on final flow\;
\caption{$O(N^5 \log N$) Baranyai construction}
\label{alg:baranyai}
\end{algorithm}

A useful aspect of the Baranyai-based approach to molecular Hamiltonian partitioning is that it depends only on $N$ and not on the $h_{pq}$ and $h_{pqrs}$ coefficients in Equation~\ref{eq:second_quantization}. In this sense, it is pre-computable---for instance, the $N=8$ partitioning in Table~I will apply to all 8-qubit Hamiltonians. By contrast, MIN-COMMUTING-PARTITION techniques in prior work operate on the specific molecular Hamiltonians of interest. Thus, the partitionings are not pre-computable and the classical cost of partitioning must be accounted for in time-to-solution.

\section{Discussion}
We have demonstrated that Jordan-Wigner encoded molecular Hamiltonians can be partitioned into $O(N^3)$ commuting families, each containing $O(N)$ Pauli strings. Our proof stems from Baranyai's Theorem, which has a constructive form that efficiently yields partitionings, per Algorithm~\ref{alg:baranyai}. Since commuting families can be measured simultaneously, this constitutes a reduction in the measurement cost of VQE from $O(N^4)$ naively to $O(N^3)$ with these partitions. The simultaneous measurement circuits are efficient too, requiring only $O(N)$ gates, since the shared eigenbasis of the commuting partitions can be expressed as a tensor product over 4-qubit chunks.

An advantage of our technique is that it only depends on $N$ and is pre-computable for all $N$-qubit molecular Hamiltonians. Further optimizations may be possible by analyzing $h_{pqrs}$ coefficients in Equation~\ref{eq:second_quantization}. For example, for molecular Hamiltonians, we expect the $h_{pqrs} = h_{srpq}$ symmetry \cite{whitfield2011simulation}, which reduces the number of relevant Pauli strings in each $\{a^{\dagger}_p a^{\dagger}_q a_r a_s\}_\text{JW}$ set from 16 to 8.

Recent work \cite{huggins2019efficient} has gone deeper in this direction, by factoring molecular Hamiltonians into a form that empirically seems to have $O(N)$ partitions. Moreover, the simultaneous measurements only appear to require $O(N^2)$ gates, even with linear qubit connectivity. It would be informative to benchmark this recent work against our strategy, which produces $O(N^3)$ partitions but requires only $O(N)$ gates under full connectivity.

Beyond VQE, our technique may be useful in other quantum computational chemistry applications. For example, the simulation of Hamiltonian \textit{dynamics} could be improved by partitioning into commuting families. Naively, Hamiltonian evolution is performed by Trotterization that requires fine time slicing to account for non-commuting terms \cite[Section 4.7]{nielsen2010quantum}. However, by ordering Pauli strings in a Hamiltonian such that large commuting sets are consecutive, the Trotterization cost could be diminished. This approach seems promising since our work proves an asymptotic gain for partitioning. Moreover, simultaneous measurement circuits would not be needed, so this re-ordering of a Trotterization would have essentially no quantum cost.

\section*{Acknowledgements}
We are grateful to our collaborators from our previous work \cite{gokhale2019minimizing}, with whom we had many helpful background conversations: Olivia Angiuli, Yongshan Ding, Kaiwen Gui, Teague Tomesh, Martin Suchara, and Margaret Martonosi. This work is funded in part by EPiQC, an NSF Expedition in Computing, under grant CCF-1730449 and in part by STAQ, under grant NSF Phy-1818914. Pranav Gokhale is supported by the Department of Defense (DoD) through the National Defense Science \& Engineering Graduate Fellowship (NDSEG) Program.
\bibliography{refs}

\end{document}